 \documentclass{camera}
\newcommand{\be}{\begin{equation}}
\newcommand{\ee}{\end{equation}}
\newcommand{\bea}{\begin{eqnarray}}
\newcommand{\eea}{\end{eqnarray}}
\newcommand{\bra}[1]{\left\langle #1 \right|}
\newcommand{\ket}[1]{\left| #1 \right\rangle}
\begin{document}
\rightline{ BARI-TH/417-2001}
\rightline{May 2001}
%
\title{HIGHLIGHTS IN THE ANALYSIS\\OF EXCLUSIVE $B$ DECAYS}

%
\author{F. De Fazio}

%
\organization{Istituto Nazionale di Fisica Nucleare - Sezione di Bari}

\maketitle

%
\abstract{I briefly describe recent developments in the theoretical
analysis of non leptonic and semileptonic  B decays.
For non leptonic transitions, I focus on factorization, from the naive
formulation to the most recent achievements.
As for semileptonic decays, I mainly consider
B transitions to excited charmed states.}

\section{Non Leptonic $B$ Meson Transitions}

The theoretical description of  non leptonic decays is very
difficult since the final state is composed only of hadrons, thus
requiring the consideration of the interplay between weak and
strong dynamics. This is achieved 
using an Operator Product Expansion to write
the effective hamiltonian describing a given weak 
decay as  a sum of local operators, weighted by Wilson coefficients,
both depending on a scale $\mu$.
The coefficients
include short distance dynamics and hence can be computed
perturbatively, while the matrix elements of the operators include
long distance physics at scales below $\mu$, 
in such a way that the dependence on the scale cancels in the product.
A  simple approach to evaluate such matrix elements
  is  the "naive" factorization \cite{naive}.
Let us consider the process ${\bar B}^0 \to D^+
\pi^-$: the effective hamiltonian is $H_{eff}=\displaystyle{G_F \over
\sqrt{2}} V_{cb}V_{ud}^*[C_1(\mu) O_1+C_2(\mu)O_2]$, where
$O_1=({\bar c}b)_{V-A}({\bar d}u)_{V-A}$, $O_2=({\bar
c}u)_{V-A}({\bar
d}b)_{V-A}$, and $({\bar q}_1 q_2)_{V-A}={\bar q}_1
\gamma_\mu (1-\gamma_5)q_2$.
Using the properties of the 
Gell-Mann matrices $T^a$, one has
$O_2=\displaystyle{O_1 \over N_C}+2{\tilde O}$,
giving
$H_{eff}=\displaystyle{G_F \over\sqrt{2}} V_{cb}V_{ud}^*\left[ a_1(\mu)
O_1 +2 C_2(\mu){\tilde O}\right]$,
with $a_1(\mu)=C_1(\mu)+C_2(\mu)/N_C$ and 
${\tilde O}=({\bar c}T^a b)_{V-A} ({\bar d}T^a u)_{V-A}$.
Naive factorization consists in factorizing the matrix element 
 $\bra{\pi^- D^+} O_1 \ket{{\bar B}^0}$ and neglecting the contribution of
${\tilde O}$. This is
because of the "colour trasparency", an argument due to Bjorken
 according to which the  quarks of the
emitted pion move fast away from the interaction region and behave as a
colour singlet, with no coupling to the coloured currents.
The final  amplitude reads as
$
{\cal A}({\bar B}^0 \to D^+ \pi^-)={G_F \over\sqrt{2}} V_{cb}V_{ud}^*
a_1(\mu) \bra{\pi^-}({\bar d}u)_{V-A}\ket{0} \bra{D^+}({\bar
c}b)_{V-A}\ket{{\bar B}^0}$
and can be expressed in terms of 
$B \to D$  form factors and $f_\pi$.
The decays taking contribution only from operators like
 $O_1$ are  referred to as Class I decays; those taking
contribution only from $O_2$ as Class II and those taking both
contributions as Class III ones. Since the matrix elements
in the factorized expression do not depend on $\mu$
any more,  the scale dependence of the $a_i$ is not matched,
therefore  naive factorization cannot be exact.

An improvement is represented by 
generalized factorization 
\cite{generalized}, describing
 non factorizable terms  through parameters
assumed universal  for processes with similar
kinematics. For example, one can write $a_1=C_1+\xi C_2$, and consider
$\xi=1/N_C$ as a free parameter to be experimentally fitted. 
Another possibility is to split a  non leptonic amplitude in 
substructures which can either be classified according to their
transformation properties under $SU(3)_F$ or associated to Wick
contractions of the operators in the effective hamiltonian
\cite{topologie}. Symmetry
arguments allow to derive relations or to establish  hierarchies among 
such substructures. 

A recent study \cite{bbns} has derived a QCD factorization formula, 
which can be
applied to many, but not all, $B$ decays in the limit $m_b \to \infty$
\footnote{An alternative approach has been proposed in \cite{keum}.}.
This analysis demonstrates that in the decay $B \to M_1
M_2$, non factorizable contributions are due to hard gluon exchange, while
soft effects are confined to the system $(B M_1)$, if $M_1$ is the meson
which picks up the spectator quark of the  $B$. 
Naive factorization is recovered at leading order in
$\Lambda_{QCD}/ m_b$ and $\alpha_s$.
The QCD factorization formula is then expressed in terms of form factors
describing the transitions $B \to M_1, M_2$ and the light cone wave
functions of the particles (we refer to \cite{bbns} for the
explicit formula).
It is proven that non factorizable 
topologies   (vertex corrections, penguin
diagrams, hard spectator interactions and annihilation diagrams) are free
from infrared divergences at leading order in
$\Lambda_{QCD}/m_b$. They 
 represent
${\cal O}(\alpha_s)$ contributions which are non universal, 
depending on the meson $M_2$ which does not pick up the spectator quark.
Finally, the absence of infrared divergences cannot be proven if $M_2$ is
a heavy meson. 

The  role of subleading
terms in $1/m_b$ remains an open question; it could be sizable if the
leading term is suppressed for some
reasons (colour suppression, CKM suppression, small Wilson coefficients).
This could be the case for $B \to K \pi$
\cite{bbns2}. The quantitative estimate of such terms would be
fundamental in many respects; in particular, it would put  the extraction
of CP violating parameters from several non leptonic
decay modes on a firmer theoretical basis.

\section{Exclusive Semileptonic $b \to c$ Processes }

In the limit $m_Q \to
\infty$, where $Q$ is a heavy quark with $m_Q \gg \Lambda_{QCD}$, 
Heavy Quark Effective Theory (HQET) 
exploits the decoupling of the light degrees of freedom 
to classify the heavy mesons in doublets \cite{hqet}. 
The members of these differ
only for the orientation of the heavy quark spin with respect to
the angular momentum of the light
degrees of freedom  ${\vec s}_\ell={\vec \ell} 
+{\vec s}_q$, $\vec \ell$
being the orbital angular momentum and ${\vec s}_q$ the light quark spin.
The low-lying doublet $(P,P^*)$, with $P=D,B$, corresponds to $\ell=0$
and has spin-parity $J^P=(0^-,1^-)$.
For $\ell=1$, the two doublets $(P_0,P_1^\prime)$ and $(P_1,P_2)$ have
$J^P=(0^+,1^+)$ and $J^P=(1^+,2^+)$, respectively.
Excited heavy mesons have
been identified in the charm sector  and observed also in the
beauty case \cite{pdg}.
The transitions between the members of two doublets are
all described in terms of a universal function: the transitions
$(P,P^*)$ $\to$ $(P,P^*)$ are described by the Isgur-Wise function
$\xi(y)$, where
$y=v \cdot v^\prime$ and $v$ and $v^\prime$ are the four velocities of
the heavy meson in the initial and final state, respectively.
$\xi$ takes the place of 6 form factors and
 is normalized to 1 in the zero recoil point. The
inclusion of short distance corrections, as well as the Luke theorem
\cite{luke}, assuring the absence of $1/m_b$ corrections in some
circumstances, lead to rather precise theoretical predictions for the
 $B \to  D^* $ semileptonic decay. The comparison with the data
allows the most accurate determination of $V_{cb}$
\cite{vcb}.
Moreover, $(P,P^*)$ $\to$ $(P_0,P_1^\prime)$ transitions
involve the universal
function $\tau_{1/2}(y)$, while $(P,P^*)$ $\to$ $(P_1,P_2)$ involve the
function $\tau_{3/2}(y)$;
the two $\tau$ functions take the place of 14
form factors.

It is  interesting to understand  the contribution of
semileptonic $B$ decays to excited charmed mesons
to the inclusive semileptonic $B$  branching ratio. Moreover, the
calculation of the
$\tau$ functions is worth due to their universality
within HQET and their role in the Bjorken and Voloshin sum rules
\cite{sr}. 
In general, HQET does not allow the determination of these
universal
functions, and, in particular, does not predict their normalization at
zero recoil:  non perturbative
techniques are required.
QCD sum rules \cite{shifman} have been widely applied to this aim
\footnote{A comparison with the results of other approaches, when
available, is provided in \cite{mia}.}.
The function $\tau_{1/2}$ has been computed
including ${\cal O}(\alpha_s )$ corrections \cite{noitau12}, which are
moderate as for the $\xi$  function \cite{neubert}, while
$\tau_{3/2 }$ has been estimated at leading order in
$\alpha_s$ \cite{lorotau32}.
The prediction for the  semileptonic branching
ratios:
\bea
BR(B \to D_0 \ell \nu_\ell)=(5 \pm 3) \times 10^{-4}  &\;&
BR(B \to D_1^\prime \ell \nu_\ell)=(7 \pm 5) \times 10^{-4} \nonumber
\\
BR(B \to D_1 \ell \nu_\ell)= 3.2 \times 10^{-3}  &\;&
BR(B \to D_2 \ell \nu_\ell)= 4.8 \times 10^{-3} \label{br}
\eea
\noindent must be compared to the  data  \cite{pdg}:
$BR(B^+ \to {\bar D}_1^0(2420) \ell^+ \nu_\ell)= (5.6 \pm 1.6) \times
10^{-3}$,
$BR(B^+ \to {\bar D}_2^{*0}(2460) \ell^+ \nu_\ell)< 8 \times 10^{-3}$,
though the ${\bar D}_1^0(2420)$ is likely to be an admixture of
the two $1^+$ states. 
The method predicts also the masses of the excited states,
with
the result: 
${\bar \Lambda}^+_{1/2}= M_{D_0}-m_c=1.0 \pm 0.1$ GeV,
${\bar \Lambda}^+_{3/2}= M_{D_1}-m_c=1.05 \pm 0.1$ GeV, 
in agreement with  existing data.

Higher states can be further considered. If $l=2$,
two doublets  $(D_1^*,D_2^*)$ and $(D_2^{\prime *},D_3^*)$
are found with $J^P_{s_\ell}=(1^-,2^-)_{3/2}$ and 
$J^P_{s_\ell}=(2^-,3^-)_{5/2}$ respectively.
QCD sum rules predict a very small 
 universal function for  $B \to (D_1^*,D_2^*)$ decays, 
so that the corresponding
transitions are negligible \cite{noi52}. However, the transitions to the
states belonging to the doublet with $s_\ell=5/2$ are presumably
observable, with predicted branching ratios
$BR(B \to D_2^{\prime *} \ell \nu_\ell) \simeq 
BR(B \to D_3^* \ell \nu_\ell) \simeq 1 \times 10^{-5}$ \cite{noi52}.
Since their strong decays  proceed through F-wave transitions, we
expect these
states to be rather narrow, which is confirmed by  the estimate of
their width obtained varying their strong coupling to pions in the range
[0.2,0.5] \cite{noi52}.

Also in this case it is important to analyse the role of  ${\cal 
O}(\displaystyle{1 \over m_b})$ corrections, when the contribution of
higher
dimensional operators in the effective currents and in the effective HQET
lagrangian should be considered \cite{wise,dai1,dai2}. 
For the semileptonic transition  
$B \to D_2 \ell \nu_\ell$ these corrections are rather modest, while they
turn
out to be large for $B \to D_1 \ell \nu_\ell$ \cite{dai2}. 
Further improvements in the analysis of such corrections, including
next-to-leading order QCD terms, are required.

\section{Conclusions}
Important theoretical progresses have been achieved in the study of
exclusive $B$
decays, mainly exploiting the large value of  $m_b$. Open
questions remain, such as the role and the classification of $1/m_b$
corrections. New
theoretical efforts are therefore required to match the 
experimental accuracy  expected in next few years.

%

\begin{thebibliography}{99}

\bibitem{naive}
J. Schwinger, Phys. Rev. Lett. {\bf 12} (1964) 630; R.P. Feynman in
Symmetries in Particle Physics, edited by A. Zichichi, 1965;
O. Haan and B. Stech, Nucl. Phys. B {\bf 22} (1970) 448

\bibitem{generalized}
M. Bauer and  B. Stech, 
Phys. Lett. B {\bf 152} (1985) 380, 
M. Bauer, B.Stech and M. Wirbel, 
Z. Phys. C {\bf 34} (1987) 103; 
A.J. Buras et al.,

Nucl. Phys. B {\bf 268} (1986) 16; 
M. Neubert et al., in Heavy Flavours,
edited by A.J. Buras and M. Lindner, (World Scientific, Singapore) 1992;
M. Neubert and B. Stech, in Heavy
Flavours II, edited by A.J. Buras and M.
Lindner, (World Scientific, Singapore) 1998

\bibitem{topologie}
D. Zeppenfeld, Z. Phys. C {\bf 8} (1981) 77; 
M.J. Savage and M.B. Wise,
Phys. Rev. D {\bf 39} (1989) 3346, {\bf 40} (1989) 3127 (E); 
L.L. Chau,
Phys. Rev. D {\bf 43} (1991) 2176; 
M. Gronau et al., Phys. Rev. Lett. {\bf 73} (1994) 21; 
Phys. Lett. B {\bf 333} (1994) 500;
Phys. Rev. D {\bf 50} (1994) 4529;
M. Ciuchini et al.,
Nucl. Phys. B {\bf 501} (1997) 271; 
Nucl. Phys. B {\bf 512} (1998) 3; {\bf 531} (1998) 656 (E); 
A.J. Buras and R. Fleischer, Phys. Lett. B 
{\bf 341} (1995) 379; 
A.J. Buras and L. Silvestrini, Nucl. Phys. B {\bf 569} (2000) 3

\bibitem{bbns}
M. Beneke, G. Buchalla, M. Neubert and C.T. Sachrajda, Phys. Rev. Lett.
{\bf 83} (1999) 1914;
 Nucl. Phys. B {\bf 591} (2000) 313

\bibitem{keum}
Y.-Y. Keum et al., Phys. Lett. {\bf B504} (2001) 6; Phys. Rev. {\bf D63}
(2001) 054008; Phys. Rev. {\bf D63} (2001) 074006.

\bibitem{bbns2}
M. Beneke et al., hep-ph/0104110;
M. Ciuchini et al., hep-ph/0104126

\bibitem{hqet} For a review see: M. Neubert, Phys. Rep. {\bf 245} (1994)
259

\bibitem{pdg}
D.E. Groom et al, Review of Particle Physics, Eur. Phys. J C {\bf 15}
(2000) 1

\bibitem{luke}
M.E. Luke, Phys. Lett. B {\bf 252} (1990) 447

\bibitem {vcb}
The LEP $V_{cb} $ Working Group, http://lepvcb.web.cern.ch/LEPVCB
 
\bibitem{sr}
J.D. Bjorken, Proceedings of the 4th Recontres de Physique de la Vallée
d'Aoste, La Tuille 1990, edited by M. Greco, (Editions Frontières, Gift
sur Yvette) 1990; 
M.B. Voloshin, Phys. Rev. D {\bf 46} (1992) 3062

\bibitem{shifman}For  reviews see:
{\it Vacuum Structure and QCD Sum Rules},
M.A. Shifman ed., North-Holland, Amsterdam, 1992;
P. Colangelo and A. Khodjamirian, hep-ph/0010175

\bibitem{mia}
F. De Fazio, hep-ph/0010007

\bibitem{noitau12}
P. Colangelo et al., Phys. Rev. D {\bf 58} (1998) 116005

\bibitem{neubert}
M. Neubert, Phys. Rev. D {\bf 47} (1993) 4063

\bibitem{lorotau32}
P. Colangelo et al., Phys. Lett. B {\bf 293} (1992) 207

\bibitem{noi52}
P. Colangelo et al., Phys. Lett. B {\bf 478} (2000) 408

\bibitem{wise}
A.K. Leibovich et al., Phys. Rev. Lett. {\bf 78} (1997) 3995; 
Phys. Rev. D {\bf 57} (1998) 308

\bibitem{dai1}
M.Q. Huang et al., Phys. Rev. D {\bf 61} (2000) 054010

\bibitem{dai2}
M.Q. Huang and Y.-B. Dai, hep-ph/0102299.

\end{thebibliography}
\end{document}